\documentclass[a4paper]{jpconf}
\usepackage{graphicx}
\usepackage{amsmath, amsfonts, amssymb}

\begin{document}
\title{Jet Multiplicity in Top-Quark Pair Events at CMS}

\author{A. Descroix for the CMS Collaboration}

\address{Karlsruhe Institut of Technologie, Wolfgang-Gaede-Str. 1, 76131 Karlsruhe, Germany}

\ead{descroix@cern.ch}

\begin{abstract}
The normalised differential top quark-antiquark production cross section is measured as a function of the jet multiplicity. Using a procedure to associate jets to decay products of the top quarks, the differential cross section of the $\mathrm{t\bar{t}}$ production is determined as a function of the additional jet multiplicity. The fraction of events with no additional jets is measured as a function of the threshold required for the transverse momentum of the additional jet. The measurements are compared with predictions from perturbative quantum chromodynamics and no significant deviations are observed.
\end{abstract}

\section{Introduction}

Top-pair events are often accompanied by additional hard jets that do not originate from the decay of the top pair ($\mathrm{t\bar{t}}$+jets) at the LHC. These events provide an essential handle to test higher-order QCD calculations of processes leading to multijet events but also to test the choice of renormalisation and factorisation scales for data modeling. The correct description of $\mathrm{t\bar{t}}$+jets production is crucial for the overall LHC physics program since it constitutes an important background for processes of interest with multijet final states. Typical processes are the associated Higgs-boson production with a $\mathrm{t\bar{t}}$ pair, where the Higgs boson decays into jets, or final states predicted in supersymmetric theories. Anomalous production of additional jets accompanying a $\mathrm{t\bar{t}}$ pair could be a sign of new physics beyond the standard model.\\Using 5.0 fb$^{-1}$ of proton-proton collisions at 7 TeV recorded in 2011 by the Compact Muon Solenoid (CMS) detector, the $\mathrm{t\bar{t}}$ cross section is measured differentially as a function of the total number of jets in the event (in the dilepton and $\ell$+jets channels), and as a function of the number of additional jets in the event (in the $\ell$+jets channel). The fraction of events that do not contain additional jets (gap fraction), is determined as a function of the threshold on the transverse momentum ($p_\text{T}$) of the leading additional jet (in the dilepton channel). A complete documentation of this analysis and all the corresponding references can be found in~\cite{us}.

\section{Event Selection}

Events in the dilepton channel are required to contain at least two isolated leptons (electrons or muons) of opposite charge and at least two jets, of which at least one is identified as a b-jet. A cut applied on events with a small lepton-pair invariant mass suppresses background from heavy-flavour resonance decays. In the $ee$ and $\mu\mu$ channels, the dilepton invariant mass is required to be outside a Z-boson mass window, and a cut on the missing transverse energy (MET) is applied. Finally a kinematic reconstruction method is used to identify the two b-jets originating from the decay of the top quark and antiquark. Those events which do not provide a valid reconstruction are rejected.\\In the $\ell$+jets channel ($\ell = e,\mu$), events are retained if they contain exactly one isolated lepton. A veto against additional leptons is applied to reject events in the dilepton decay channel. The presence of at least three reconstructed jets is required and at least two of them must be b-tagged.\\A particle-level selection defines the phase space of the results. This selection is solely based on simulated particles before applying the detector simulation. The kinematic requirements for the charged lepton(s) originating from the $\mathrm{t\bar{t}}$ decay and for the jets are similar to those at the detector-level. Jets are identified as originating from b-quarks if they contain the decay products of a B hadron.

\section{Signal and Background Modeling}

The signal is simulated by {\sc MadGraph} (v.\,5.1.1.0) interfaced with {\sc pythia} (v.\,6.424), with $\mathrm{t\bar{t}}$+0/1/2/3-jets processes being generated at matrix-element level. Data-driven techniques are used for the modelling of $Z/\gamma^{*}$+jets in the dilepton channel and $W$+jets and QCD multijet in the $\ell$+jets channel. The remaining backgrounds are estimated by MC simulation.\\In the $ee$ and $\mu\mu$ channels, the normalisation of the $Z/\gamma^*$+jets production is derived from the data events rejected by the Z-boson veto, scaled by the ratio of events failing and passing this selection estimated in simulation.\\In the $\ell$+jets channel a QCD-multijet model is extracted from data by selecting non-isolated leptons. The normalisation is estimated with a template fit of MET using events with one b-tagged jet. The normalisation of the $W$+jets simulation in the $\ell$+jets channel is corrected, making use of the charge-asymmetrical $W$-boson production in pp collisions. In addition, the differences between data and simulation observed in the heavy-flavour content are corrected in the $W$+jets simulation.

\section{Jet- and Additional-Jet-Multiplicity Measurement}

\begin{figure*}[htb!]
  \centering
     \includegraphics[width=0.38 \textwidth]{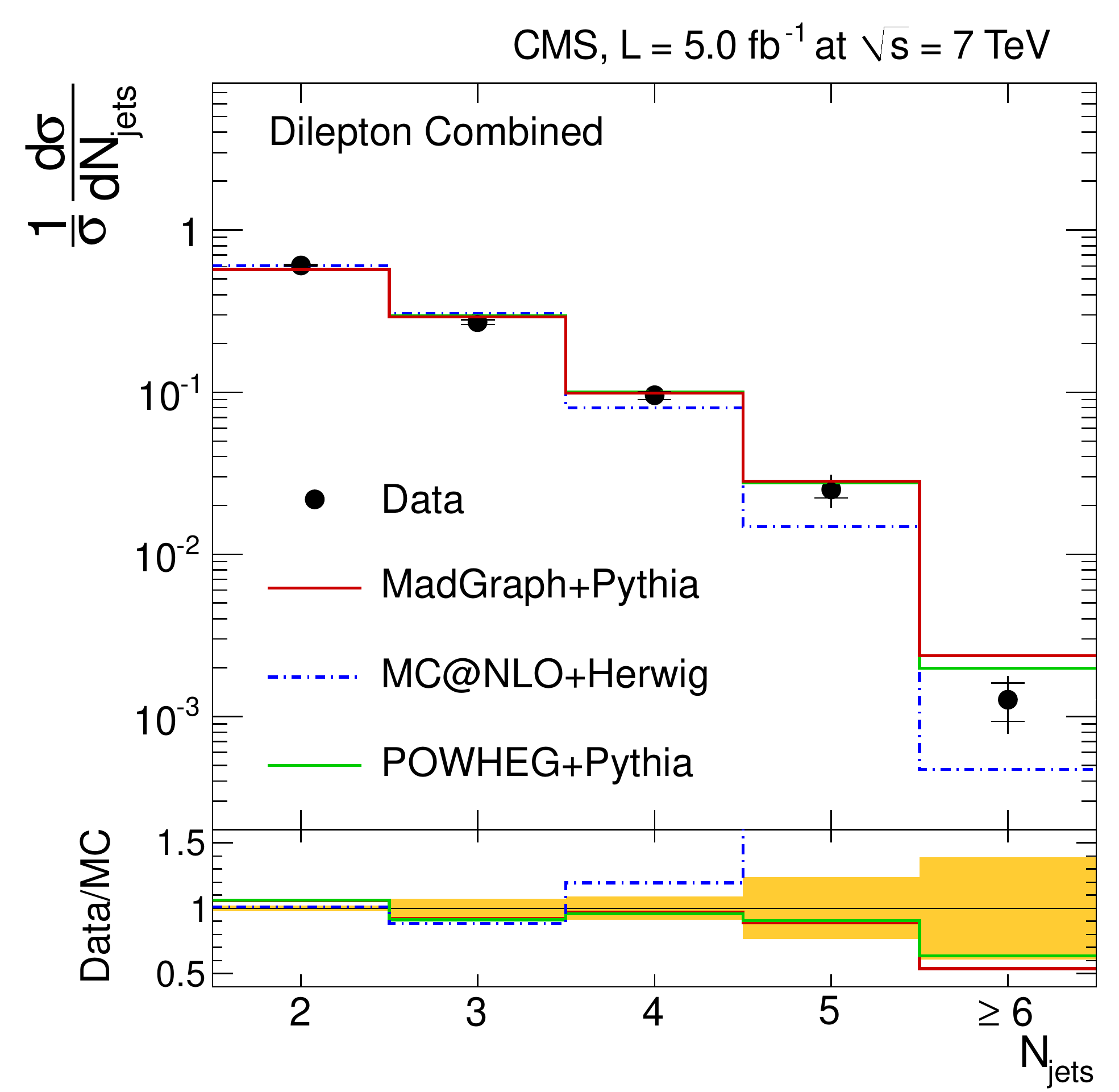}
     \includegraphics[width=0.38 \textwidth]{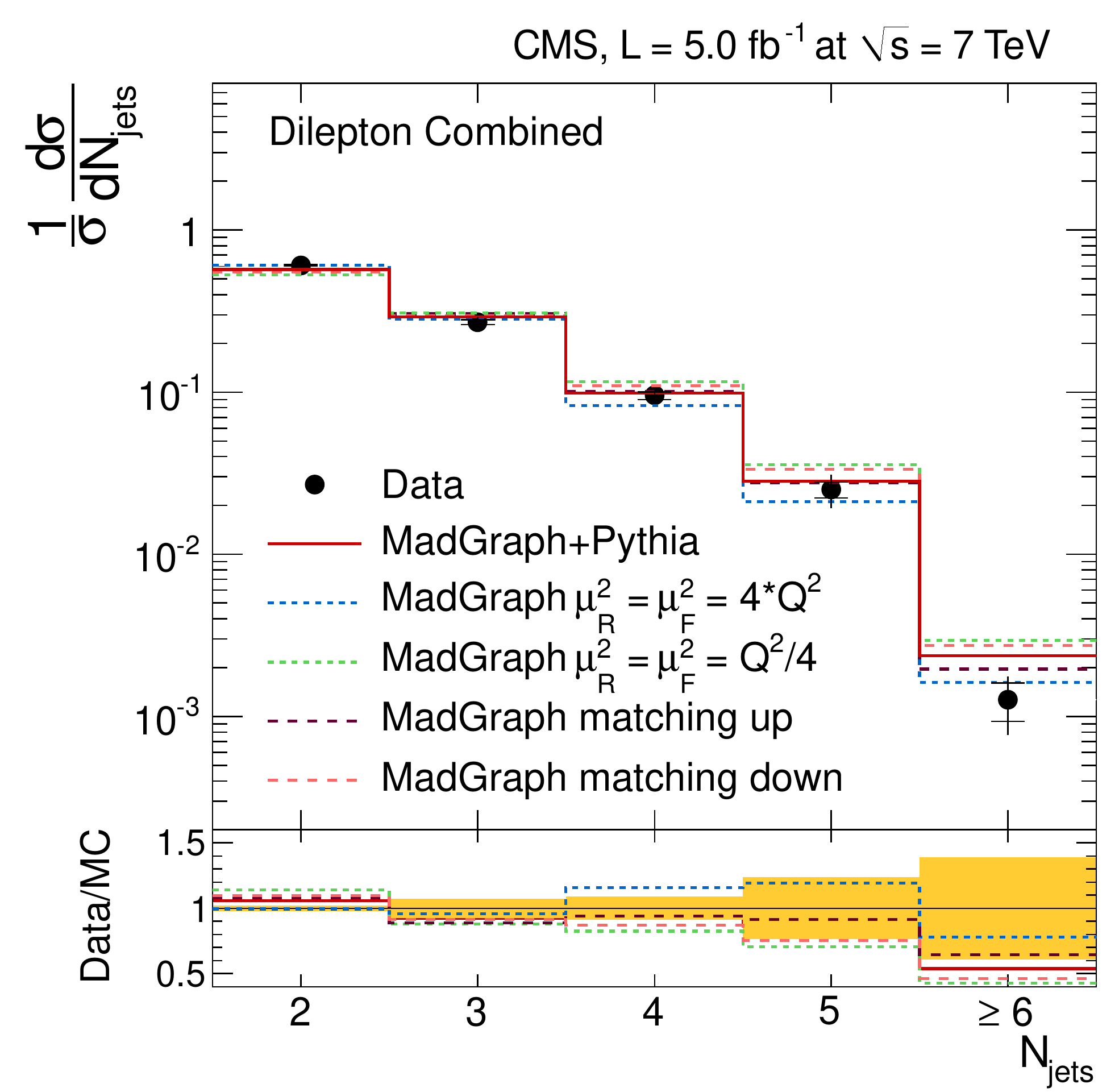}
     \includegraphics[width=0.38 \textwidth]{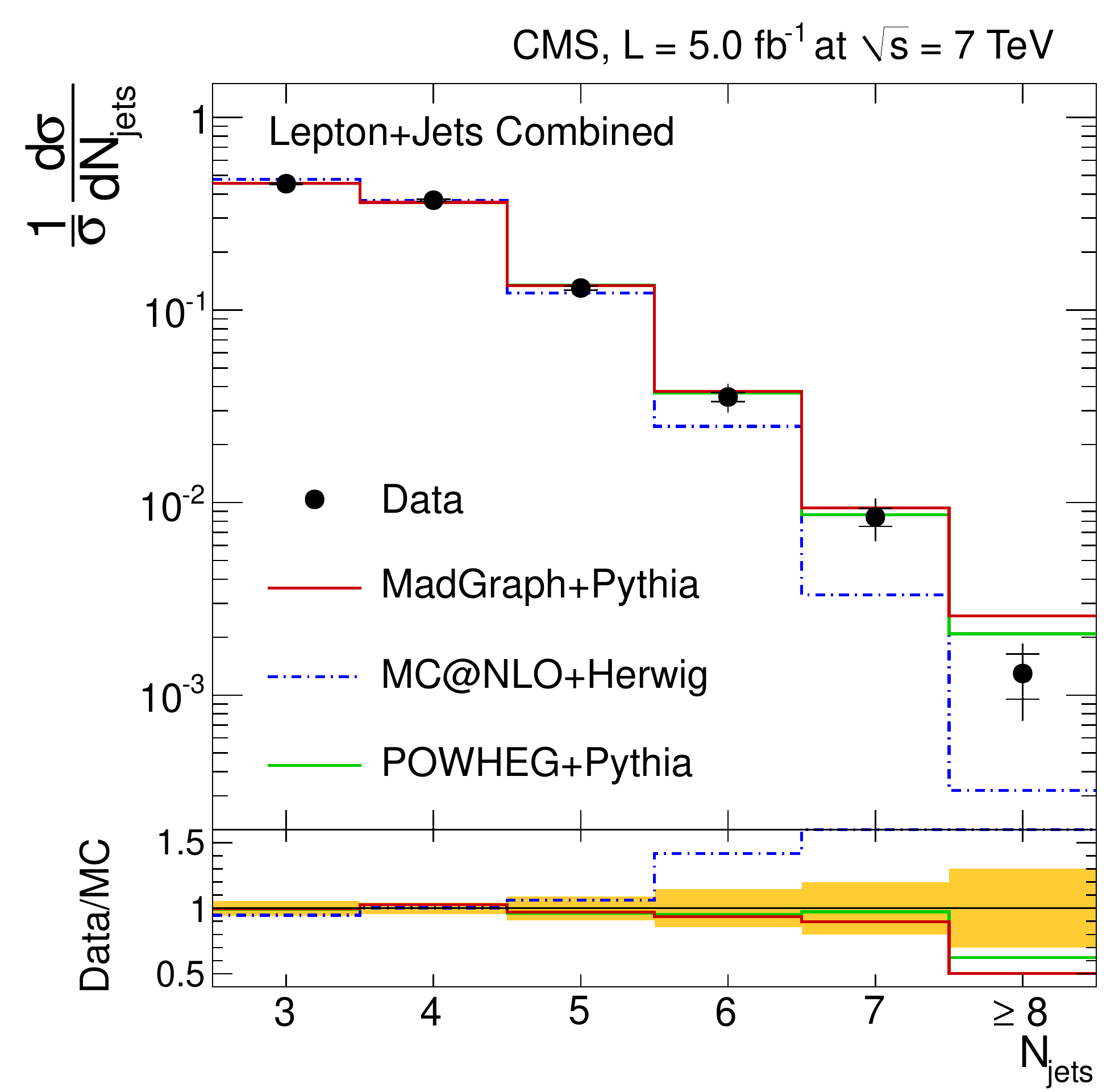}
     \includegraphics[width=0.38 \textwidth]{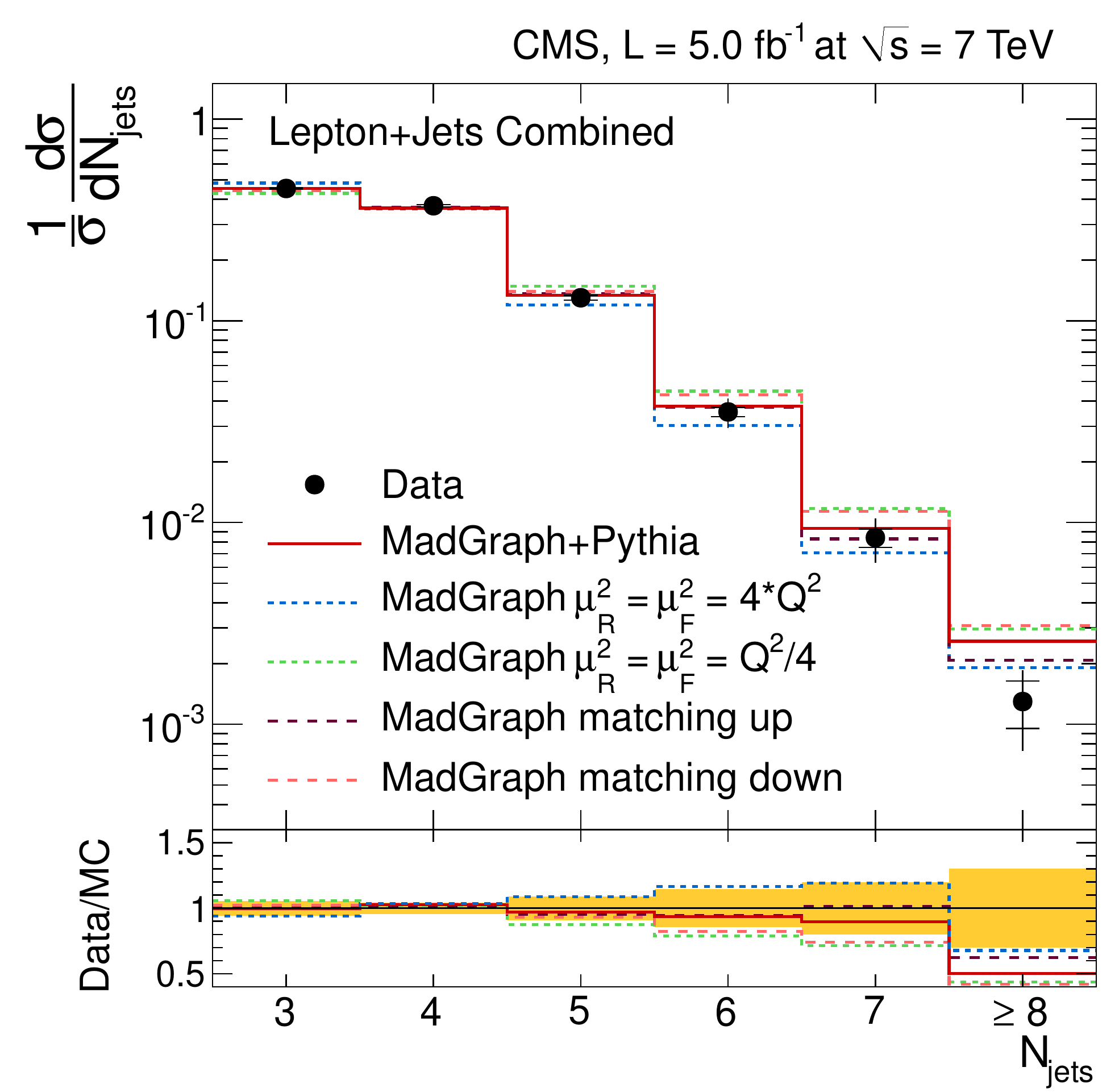}
     \includegraphics[width=0.38 \textwidth]{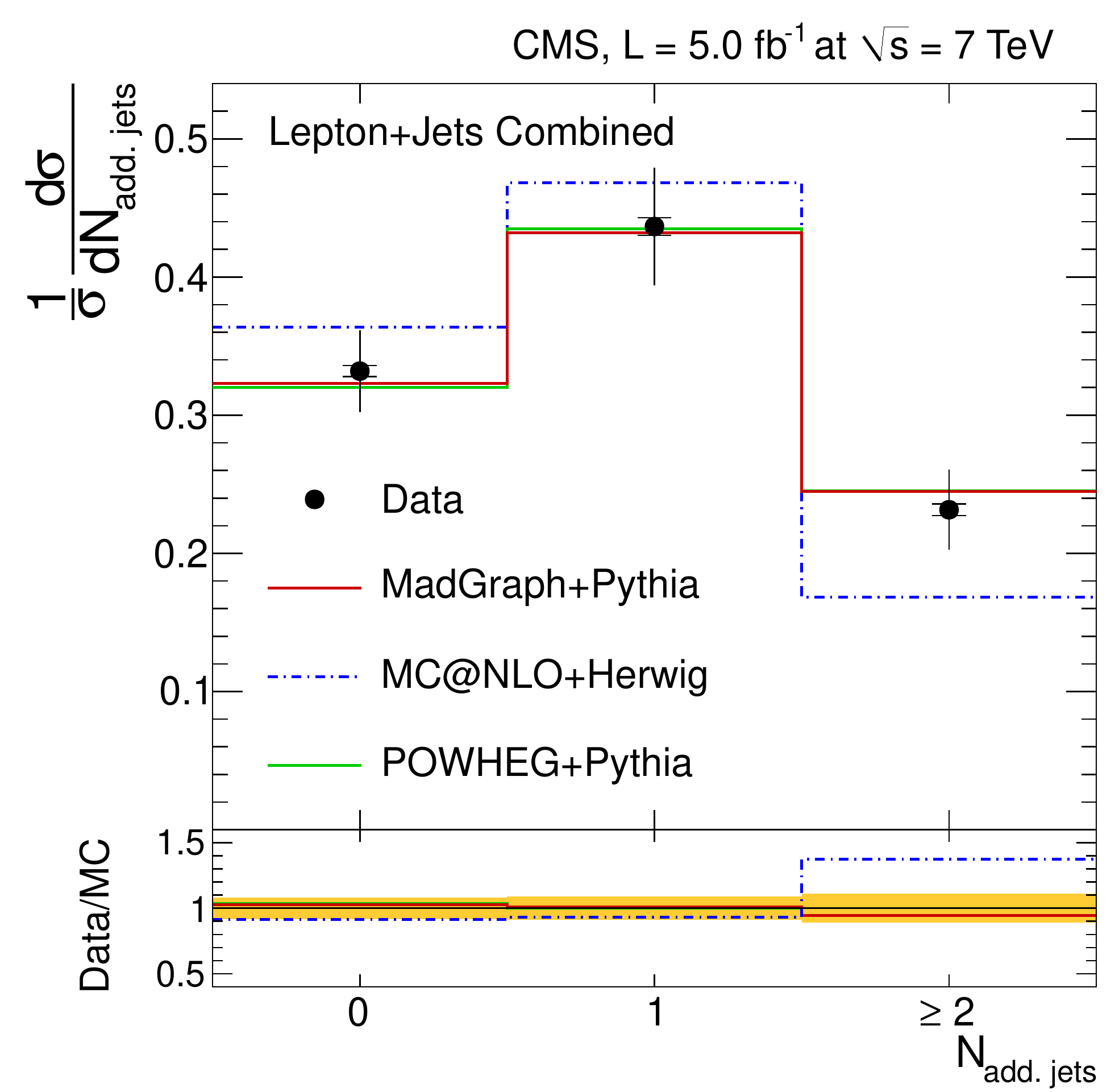}
     \includegraphics[width=0.38 \textwidth]{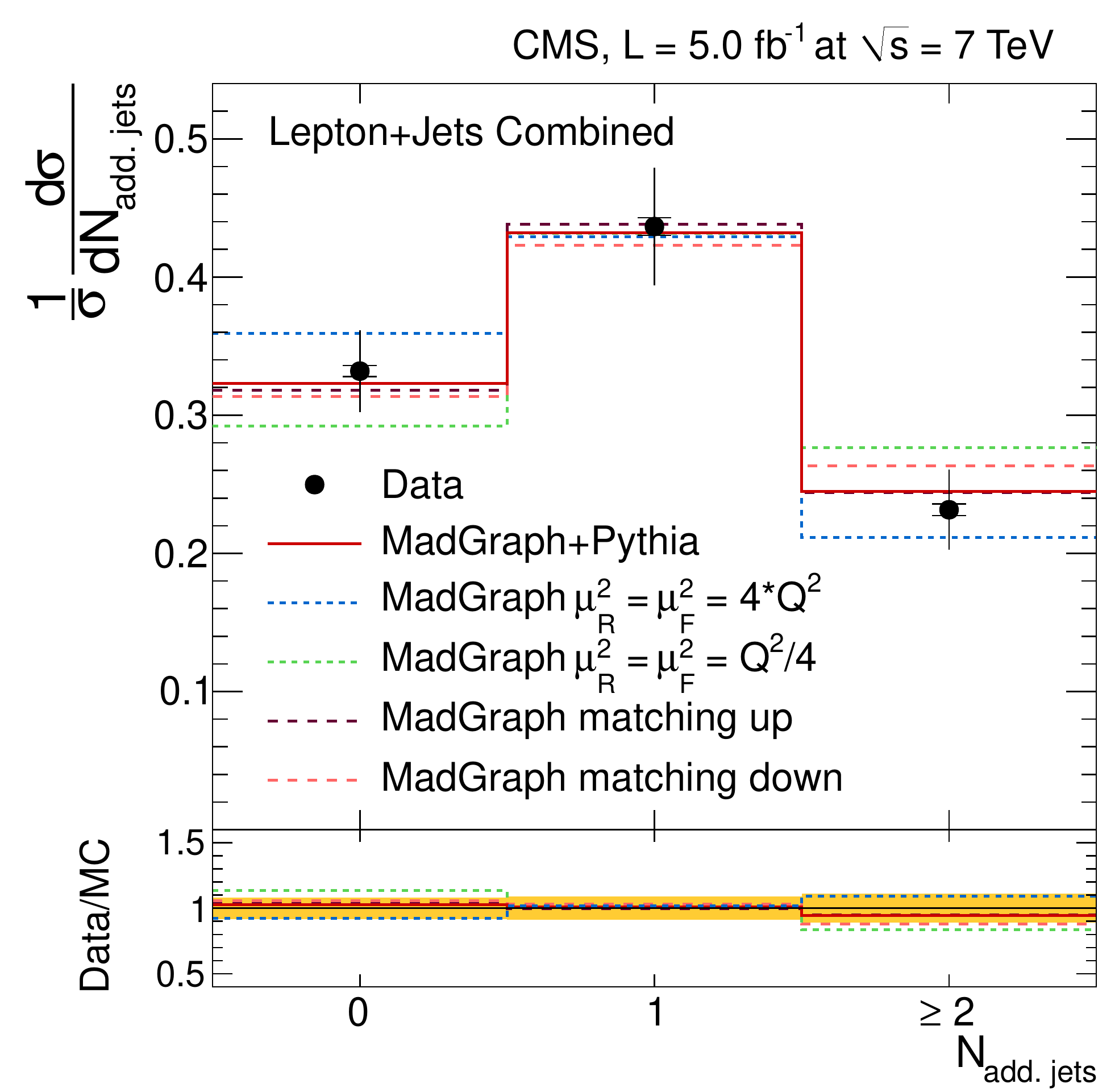}
\caption{Normalised differential $\mathrm{t\bar{t}}$ production cross section as a function of the jet multiplicity in the dilepton channel (top), in the $\ell$+jets channel (middle), and as a function of the number of additional jets in the $\ell$+jets channel (bottom). The measurements are compared to predictions from {\sc MadGraph+pythia}, {\sc powheg+pythia}, and {\sc mc@nlo+herwig} (left), as well as from {\sc MadGraph} with varied renormalisation and factorisation scales and jet-parton matching threshold (right). The inner (outer) error bars indicate the statistical (combined statistical and systematic) uncertainty. The shaded band corresponds to the combined statistical and systematic uncertainty.\label{fig:xsecjet}}
\end{figure*}

After the full selection and the subtraction of background contributions, the jet multiplicity spectrum in data is unfolded by means of a regularised single-value-decomposition method~\cite{bib:svd,bib:blobel}, in order to correct for detector-induced migration effects. This step provides results within the particle-level phase space. The unfolded spectrum is normalised by the measured cross section, in order to eliminate uncertainties stemming from normalisation effects, which leads to the results presented in Fig.~\ref{fig:xsecjet}.\\Additionally, the normalised differential cross section of $\mathrm{t\bar{t}}$ production as a function of the number of additional jets is measured in the $\ell$+jets channel, requiring $\geq$ 4 jets. Additional jets are defined as particle-level jets which do not match the direction of any of the $\mathrm{t\bar{t}}$ decay partons (the two light jets and the lepton from the $W$-boson decays and the two b-jets from the top decays). Simulated $\mathrm{t\bar{t}}$ events are categorised in $\mathrm{t\bar{t}}$+0, 1, and $\geq$2 additional jets depending on how many non-matching particle-level jets have been identified. A template fit is performed in order to extract the differential cross sections for $\mathrm{t\bar{t}}$+0, 1, and $\geq$2 additional jets categories from data. Finally the resulting cross sections are normalised and given in Fig.~\ref{fig:xsecjet}. The main systematic uncertainties are due to uncertainties on the jet energy and the signal modeling.\\Both differential cross-section measurements lead to similar conclusions: The predictions from {\sc MadGraph+pythia} and {\sc powheg+pythia} are found to provide a reasonable description of the data. In contrast, {\sc mc@nlo+herwig} generates fewer events in bins with large jet multiplicities. The comparison of the measurements with predictions from {\sc MadGraph+pythia} with variation of the renormalisation and factorisation scales and of the jet-parton matching threshold shows for both parameters a worse agreement when a lower value is chosen.

\section{Gap-Fraction Measurement}

\begin{figure*}[htb!]
  \centering
      \includegraphics[width=0.38 \textwidth]{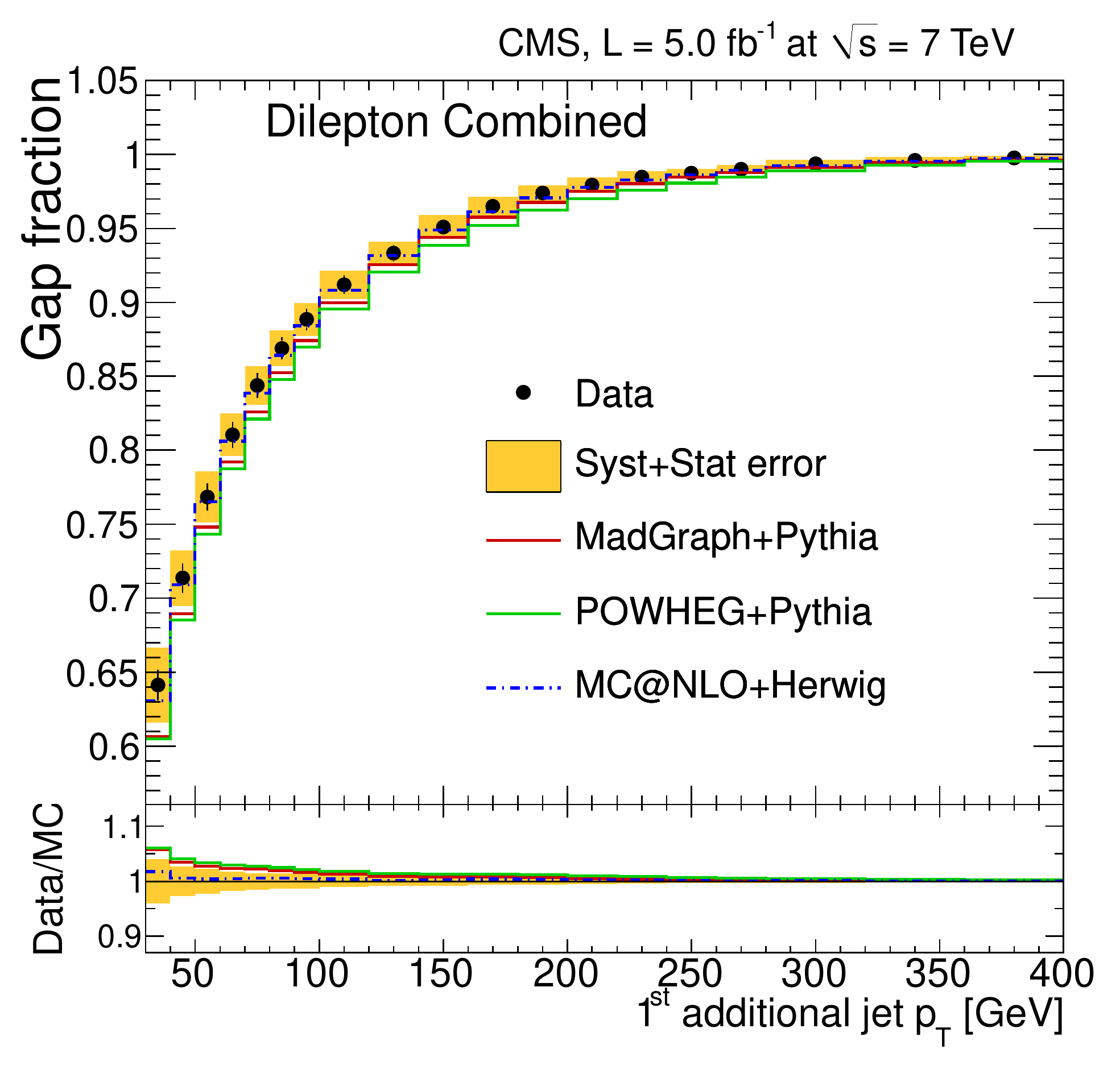}
      \includegraphics[width=0.38 \textwidth]{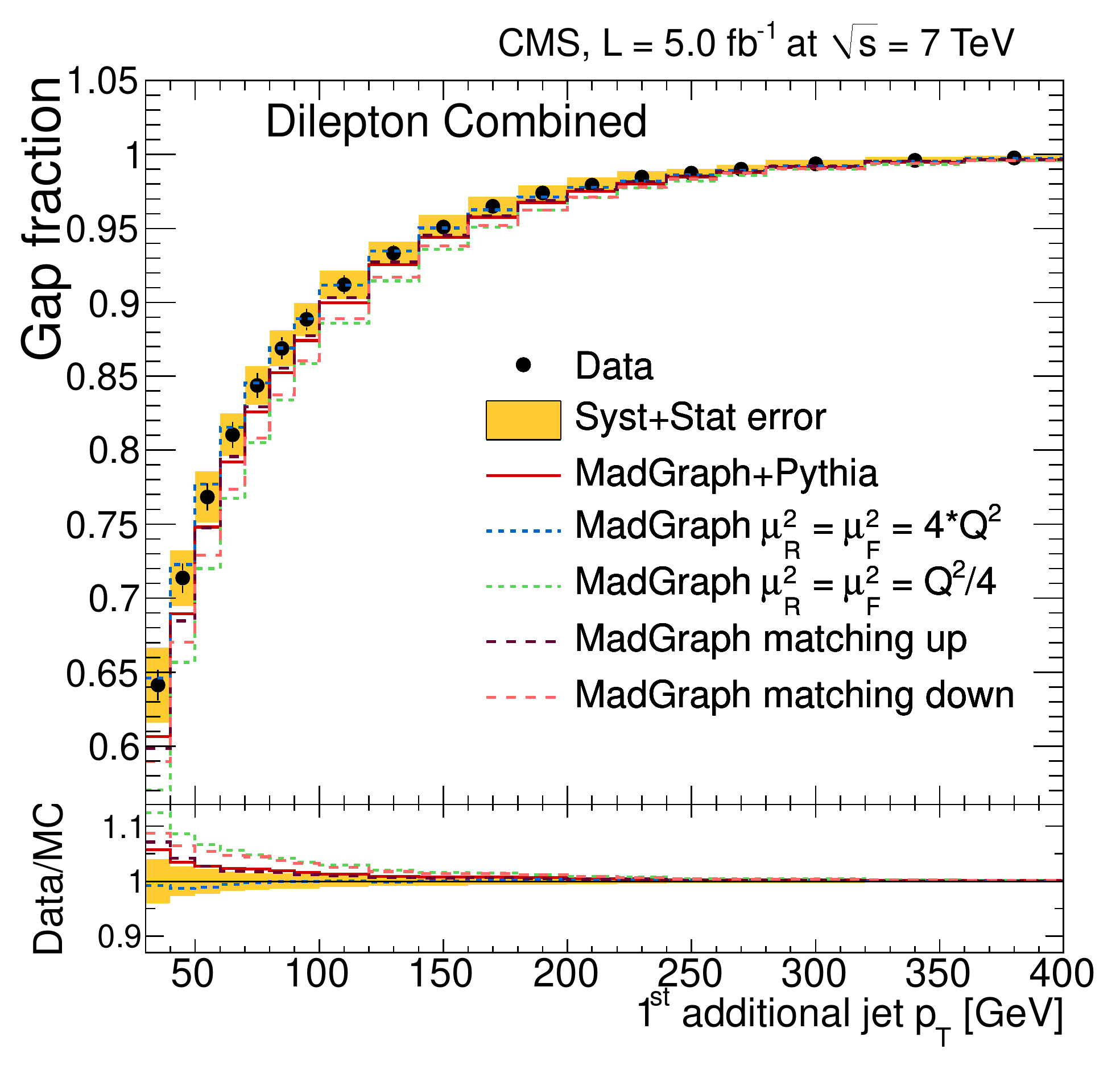}
\caption{Measured gap fraction as a function of the additional jet $p_\text{T}$ in the dilepton channel. The data is compared to predictions from the same simulated samples and the uncertainties represented the same way as in Fig.~\ref{fig:xsecjet}.\label{fig:gap_dilep}}
\end{figure*}

The jet activity arising from quark and gluon radiation in association with the $\mathrm{t\bar{t}}$ system is investigated with the fraction of events that do not contain additional jets above a given $p_\text{T}$ threshold in the dilepton decay channel. The additional jets are defined as those not assigned to the $\mathrm{t\bar{t}}$ system by the kinematic reconstruction. A bin-by-bin correction for detector effects is applied on the calculated gap fraction values. The results are shown in Fig.~\ref{fig:gap_dilep}. The main systematic uncertainties originate from the jet-energy uncertainty and the background contaminations. A better agreement between data and simulation is observed for {\sc mc@nlo+herwig} compared to {\sc MadGraph+pythia} and {\sc powheg+pythia}. This result is not incompatible with the observation described in the previous section because the gap fraction tests the description of the $p_\text{T}$ spectrum of the hardest additional jet which does not imply a good description of the larger jet multiplicity. Decreasing the renormalisation and factorisation scales or the jet-parton matching threshold in the {\sc MadGraph} sample worsens the agreement between data and simulation.

\section{Summary}

Measurements of differential cross sections with respect to jets accompanying $\mathrm{t\bar{t}}$ events are performed using a data sample corresponding to an integrated luminosity of 5.0 fb$^{-1}$ collected in pp collisions at $\sqrt{s} = 7$ TeV with the CMS detector. The results are presented in the visible phase space after correction for detector effects and compared with predictions from perturbative quantum chromodynamics from {\sc MadGraph} and {\sc powheg} interfaced with {\sc pythia}, and {\sc mc@nlo} interfaced with {\sc herwig}, as well as {\sc MadGraph} with varied renormalisation and factorisation scales, and jet-parton matching threshold. The normalised differential $\mathrm{t\bar{t}}$ production cross section is measured as a function of the number of jets in the dilepton and $\ell$+jets channels and as a function of the number of jets radiated in addition to the $\mathrm{t\bar{t}}$ decay products in the $\ell$+jets channel. The {\sc MadGraph+pythia} and {\sc powheg+pythia} predictions describe the data well up to high jet multiplicities, while {\sc mc@nlo+herwig} predicts fewer events with large number of jets. The gap fraction is measured in dilepton events as a function of the $p_\text{T}$ of the leading additional jet and is compared to different theoretical predictions. No significant deviations are observed between data and simulation. The {\sc mc@nlo+herwig} model seems to describe the gap fraction for all values of the thresholds more accurately compared to {\sc MadGraph+pythia} and {\sc powheg+pythia}. In all measurements a higher parameter value chosen for the renormalisation and factorisation scale or for the jet-parton matching threshold improves the modeling of data.

\end{document}